\documentstyle[epsfig]{mn}

\begin{document}
\def\ltsima{$\; \buildrel < \over \sim \;$}
\def\simlt{\lower.5ex\hbox{\ltsima}}
\def\gtsima{$\; \buildrel > \over \sim \;$}
\def\simgt{\lower.5ex\hbox{\gtsima}}

\title[The X-ray spectra of Compton-thick Seyfert~2 galaxies]
{The X-ray spectra of Compton-thick Seyfert~2 galaxies as seen by 
BeppoSAX}

\author[G. Matt, et al.]
{G. Matt$^{1,2}$, A.C. Fabian$^2$, M. Guainazzi$^3$, K. Iwasawa$^2$, L. 
Bassani$^4$ and G. Malaguti$^4$\\
$^1$ Dipartimento di Fisica ``E.Amaldi", Universit\'a degli Studi Roma 
Tre, Via della Vasca Navale 84, I-00146 Roma, Italy \\
$^2$Institute of Astronomy, University of Cambridge, Madingley Road, Cambridge 
CB3 0HA \\
$^3$XMM SOC, VILSPA--ESA, Apartado 50727, E--28080 Madrid, Spain\\
$^4$Istituto Tecnologie e Studio delle Radiazioni Extraterrestri, C.N.R., Via 
Gobetti 101, I-40129 Bologna, Italy \\
}

\maketitle
\begin{abstract}
Results from BeppoSAX observations of Compton--thick Seyfert 2
galaxies are summarized and reviewed, and their general properties
derived and discussed. In five out of the seven observed sources, the
nucleus is directly visible at high X-ray energies, where the photons
penetrate absorbers with column densities in the range
1.1--4.3$\times10^{24}$ cm$^{-2}$ (in the other two sources, NGC~1068
and NGC~7674, the nucleus is instead totally obscured at all energies,
implying even larger column densities). In most sources there is
unambiguous evidence of a reflection component from optically thick,
cold matter, while in two (or maybe four) cases there is also evidence
of reflection from ionized matter. For the sources with a measured
X--ray luminosity, a comparison with the infrared luminosity is made;
while in two cases (the Circinus galaxy and NGC~4945) the IR emission
appears to be dominated by starburst activity, in the other three
sources (NGC~6240, Mrk~3 and TOL~0109-383) it is likely to be
dominated by reprocessing of the UV and X--ray photons emitted by an
AGN.

\end{abstract}


\section{Introduction}

Compton--thick Seyfert 2 galaxies are by definition those AGN in which
the X--ray obscuring matter has a column density equal to or larger
than the inverse Thomson cross section, i.e. $N_H \geq \sigma_T^{-1} =
1.5\times10^{24}$ cm$^{-2}$. The Thomson cross section is equal to the
photoelectric cross section at around 10 keV (assuming cosmic
abundances), and this energy may be assumed as the boundary between
photoelectric-- and Compton--dominated regimes. By chance, this is
also the upper energy of the working band of many past X--ray
satellites, which therefore could observe Compton--thick sources only
in the photoelectric regime, where the X--ray emission is dominated by
scattered components. BeppoSAX (Boella et al. 1997), thanks to the
unprecedented sensitivity of its high energy collimated detector, the
PDS (Frontera et al. 1997), has now extended the sensitive observing
range well into the Compton--dominated regime.

There are several reasons why Compton--thick sources deserve to be
studied. Firstly, most AGN, in the local universe at least, are
obscured by Compton--thick matter (Maiolino et al. 1998). Therefore,
they are an important ingredient not only of the Cosmic X--ray
Background, but also of the IR background, where most of the absorbed
radiation is re--emitted (Fabian \& Iwasawa 1999). Secondly, the heavy
absorption means that spectral components, which would otherwise have
been completely dominated by the nuclear emission, can be observed. In
particular, in the $\sim$1--10 keV band the emission is dominated by
reflection from both cold and ionized matter of the nuclear radiation,
and the geometrical and physical properties of the circumnuclear
matter can then be studied (e.g. Matt, Brandt \& Fabian 1996).

In this paper we summarize and discuss the results from the BeppoSAX
Core Program on bright Compton--thick Seyfert 2 galaxies, as well as
sources observed in other programs which were found to be
Compton--thick, and explore the consequences.


\begin{table*}
\caption{Exposure times and count rates for the  BeppoSAX observations
of Compton--thick sources.} 
\begin{tabular}{||l|c|c|c|c|c||}
\hline
& & & & & \cr
Source &   Date~of~obs & Exp.~time~(MECS) & LECS~CR & MECS~CR & PDS~CR \cr
&   & (ks) & (s$^{-1}$) & (s$^{-1}$) & (s$^{-1}$) \cr
& & & & & \cr
  NGC~1068$^{1,2}$ & 1996-Dec/1998-Jan-11 & 101.6/37.3 & 0.110 & 
0.096$^{a}$/0.073$^{b}$  & 0.21 \cr 
  Circinus Galaxy$^{2,3}$ & 1998-Mar-13  & 137.7 & 0.065 & 0.132$^{b}$  & 2.01 \cr 
  NGC~6240$^4$ & 1998-Aug-14  & 119.4 & 0.012 & 0.024$^{b}$  & 0.38 \cr 
  Mrk~3$^{5}$ & 1997-Apr-16 & 112.8 & 0.021  & 0.069$^{a}$  & 1.16 \cr 
  NGC~7674$^{6}$ & 1996-Nov-25 & 116.0 & 0.003 & 0.005$^{a}$ & 0.13 \cr 
  NGC~4945$^7$ & 1999-Jul-01 & 93.8 & 0.031 & 0.057$^{b}$  & 2.77 \cr 
  TOL~0109-383$^8$ & 1999-Jul-26 & 64.3 & 0.005 & 0.010$^{b}$  & 0.16 \cr 
& & & & & \cr
\hline
\end{tabular}
~\par
1) Matt et al. 1997; 2) Guainazzi et al. 1999; 3) Matt et al. 1999; 
4) Vignati et al. 1999; 5) Cappi et al. 1999;  
6) Malaguti et al. 1998; 7) Guainazzi et al. 2000a; 8) Iwasawa et al. 2000;\par
$^{a}$3 MECS units; $^{b}$2 MECS units (see Sec.2)
\end{table*}

$H_0$=50 km s$^{-1}$ Mpc$^{-1}$ is adopted throughout the paper.

\section{BeppoSAX observations and results on single sources}

In this section we recall the main BeppoSAX results on NGC~1068 (Matt
et al. 1997, Guainazzi et al. 1999), the Circinus Galaxy (Matt et al.
1999, Guainazzi et al. 1999), NGC~6240 (Vignati et al. 1999), Mrk~3
(Cappi et al. 1999), NGC~7674 (Malaguti et al. 1998), NGC~4945
(Guainazzi et al. 2000a), and Tololo~0109-383 (Iwasawa et al. 2000).
In the following we will make use of results from three BeppoSAX
instruments: the LECS, MECS and PDS, working in the 0.1-10, 2-10 and
15-200 energy ranges, respectively. Details on the data reduction and
analysis can be found in the above papers.

A summary of the observations is given in Table~1, and 
the main results of the spectral fittings in Table~2. On 1997 May 6
one of the 3 MECS units failed, and therefore observations after that date
have been performed with 2 MECS units only.

\subsection{NGC~1068}

The archetypal Seyfert 2 and  Compton--thick source NGC~1068 has been 
observed by BeppoSAX
twice, one year apart. The nucleus is completely obscured at all energies 
(Matt et al. 1997), and therefore
the column density of the absorbing matter should exceed $\sim$10$^{25}$ 
cm$^{-2}$. The soft X--ray band is dominated by thermal--like emission, 
probably related to the starburst region. 
In the 2--10 keV band, the emission is a mixture of cold and ionized 
reflection (see also Iwasawa, Fabian \& Matt 1997), the latter
being complex as implied by the line spectrum (Netzer \& Turner 1997; 
Guainazzi et al. 1999). At higher energies, is the cold 
reflection component which dominates. 

There is also evidence for energy--dependent flux variability 
(Guainazzi et al. 2000b),
best explained by a variation of the spectral shape of the ionized reflector
component, obviously echoing a variation in the primary, nuclear continuum. 
This would limit the dimension of the reflecting region to less than 1 pc.

\subsection{The Circinus Galaxy}

At variance with NGC~1068, is the Circinus Galaxy the nucleus of which can be 
directly seen in the PDS band, piercing through a 4.3$\times$10$^{24}$ 
cm$^{-2}$ absorber. Here and after the transmission components have been 
modeled including Compton scattering from a spherical distribution of 
matter (an assumption which will be justified in Sec. 3.1)
as described in Matt et al. (1999) and 
Matt, Pompilio \& La Franca (1999). Cold reflection 
is evident, as both K$\alpha$ and K$\beta$ iron  fluorescent lines 
have been clearly detected. On the contrary, no clear evidence of ionized 
reflection is present, the Mg, Si and S lines (Matt et al. 1996; 
Guainazzi et al. 1999) possibly arising from the same medium producing the
above iron lines (Bianchi et al. 2000). The 
intrinsic 2--10  keV luminosity deduced from the best fit 
is $\sim$10$^{42}$ erg s$^{-1}$.

\subsection{NGC~6240}

The direct nuclear emission of NGC~6240 is revealed by the PDS (Vignati
et al. 1999; see also Netzer, Turner \& George 1998 for ASCA results). The
column density is ~2$\times10^{24}$ cm$^{-2}$. An ionized reflector is also
clearly present: a power law continuum is required by the data, and 
there is clear evidence for ionized iron lines (see also Iwasawa \& Comastri
1998). Evidence for cold reflection is instead ambiguous. The 
intrinsic 2--10 keV luminosity deduced from the best fit 
is $\sim$1.2$\times10^{44}$ erg s$^{-1}$.

\subsection{Mrk~3}

The BeppoSAX data on this source have been discussed in detail by Cappi
et al. (1999). For an easier comparison with the other sources
in this sample, we re-analyzed the high energy part of the spectrum 
adopting the transmission model of Matt, Pompilio \& La Franca (1999). 
The direct nuclear radiation is seen through an absorber of
1.1$\times$10$^{24}$ cm$^{-2}$ (and therefore, strictly speaking,
the source is not Compton--thick according to the above definition). 
The intrinsic 2--10 keV luminosity is $\sim$0.9$\times$10$^{44}$
erg s$^{-1}$. 

A cold 
reflection component is clearly required by the data (Cappi et al. 1999).
There is evidence for an unabsorbed power law component too,
suggesting the presence of an ionized reflector.  Moreover,
the iron line is broad, which would suggest a blend of cold and
ionized lines. However, ASCA did not find evidence for a substantial
broadening of the line, even if a weak additional line at $\sim$6.8 keV is 
possibly present (Iwasawa 1996). The existence of an ionized reflector must 
therefore still be considered an open issue. 

\subsection{NGC~7674}

This source has been analised and discussed by Malaguti et al. (1998). 
As for NGC~1068, the nucleus is completely hidden at all energies, 
which implies a column density $\simgt$10$^{25}$ cm$^{-2}$. The emission 
above a few keV is dominated by a cold reflection component, and the 
evidence for ionized reflection is scanty. The apparently broad iron line can 
actually be fitted by a blend of K$\alpha$ and K$\beta$ fluorescent lines. 

\subsection{NGC~4945}

This is the archetypal ``moderately thick" source, where the 
nucleus becomes visible above about 10 keV (Iwasawa et al. 1993; Done,
Smith \& Madejski 1996). Modeling the BeppoSAX data (Guainazzi et al. 2000a)
with the transmission model mentioned before, a column density for the
absorber of 2.2($^{+0.3}_{-0.4}$)$\times$10$^{24}$ cm$^{-2}$, and a power
law photon index of 1.4$\pm$0.3 are obtained. 
The intrinsic 2--10 keV luminosity is  3$\times$10$^{42}$ erg s$^{-1}$.
This value differs significantly from that quoted in Iwasawa et al. (1993)
because they did not include Compton scattering in the fitting model. 

Contrary to the other sources, no clear evidence is found for either
the cold or the ionized reflector, the spectrum below 8 keV being
dominated by extended, rather than nuclear, emission.

\subsection{TOLOLO~0109-383}

Tololo~0109-383 (NGC~424) has been observed by BeppoSAX on 1999 July 26--28
for a net exposure time of 64 ks in the MECS and PDS, and 24.6 ks
in the LECS. The fit with the usual model gives a column density 
of about 2$\times$10$^{24}$ cm$^{-2}$ (with the power law index fixed
to 2: the quality of the data is not good enough to allow for a 
simultaneous estimate of both the column density and the power law
index). The resulting intrinsic 2--10 keV luminosity is 2$\times$10$^{43}$ erg
s$^{-1}$. Evidence for both a cold and ionized reflection components
are present. 


A complete analysis of the BeppoSAX data, along with those
from a previous ASCA observation, will be presented in Iwasawa et al. (2000).



\begin{table}
\caption{Summary of the main properties of the sources in the sample. CR and
WR stand for Cold reflection and Warm reflection.}
\begin{tabular}{||l|c|c|c|c||}
\hline
& & & & \cr
Source &   $N_H^a$ &   CR &   WR & 
  $L_X^b$ \cr
 & & & & \cr
  NGC~1068$^{1,2}$ &   $\simgt$10 &   Y &   Y &   ?~($>1$) \cr
  Circinus Galaxy$^{2,3}$ &   4.3 &   Y &   N~(?) &   $\sim$0.01 \cr
  NGC~6240$^4$ &   2.2 &   Y~(?) &   Y &   $\sim$1.2 \cr
  Mrk~3$^{5,9}$ &  1.1 &  Y & Y?  & 0.9  \cr
  NGC~7674$^{6}$ &  $\simgt$10   & Y & N~(?)  & ?  \cr
  NGC~4945$^7$ &  2.2 &  N & N & $\sim$0.03  \cr
  TOL~0109-383$^8$ &  2.0 &  Y &   Y? &   $\sim$0.2 \cr
& & & & \cr
\hline
\end{tabular}
~\par
$^a$in units of 10$^{24}$~cm$^{-2}$; $^b$ 2--10 keV luminosity 
in units of 10$^{44}$~erg s$^{-1}$.\par
1) Matt et al. 1997; 2) Guainazzi et al. 1999; 3) Matt et al. 1999; 
4) Vignati et al. 
1999; 5) Cappi et al. 1999;  6) Malaguti et al. 1998; 7) Guainazzi et al.
2000a; 8) Iwasawa et al. 2000; 9) this paper
\end{table}

\section{General comments on highly obscured sources}

\subsection{The covering factor}
The first question to address is the spatial distribution of the X--ray
absorbing matter, and in particular its covering factor. Even
if formally the transmitted spectrum depends on the covering factor,
as it changes the importance of Compton
scattering into our line of sight of photons initially emitted in other
directions (Matt, Pompilio \& LaFranca 1999), the  quality of the data 
is not good enough to reach any conclusion from spectral fitting. 
We have therefore to resort to indirect arguments.  The first one 
comes from the common presence of reflection from cold matter, signaled
by the 6.4 keV fluorescent iron line and the Compton reflection component,
the latter often dominating the spectrum at energies where the nuclear
emission is completely blocked. When the intensities of these
features are compared with the direct continuum, 
a rather large covering factor is deduced (see next section). 

Further, and even stronger, evidence in favour of a large covering
factor comes from a statistical argument. As shown by Maiolino et al. (1998)
and Risaliti, Maiolino \& Salvati (1999), the fraction of Compton--thick
sources is fairly high, being possibly as large as one half
of all Seyfert 2 galaxies.
(Because the Risaliti, Maiolino \& Salvati 1999 sample is based on [OIII]
fluxes, this fraction may even be underestimated, as e.g. NGC~6240 and
NGC~4945 would have been missed, see below). 
The covering factor of the Compton--thick matter must, therefore, be large.


\subsection{How common are they?}

As mentioned above, the fraction of Compton--thick sources among
optically-selected Seyfert 2 galaxies is large, 50 per cent at least.
As Seyfert 2 galaxies outnumber Seyfert 1 galaxies by a large factor,
this means that most AGN are heavily obscured. Moreover, this fraction
may be even larger among IR--selected sources. A simple argument may
be used for a crude estimate of the fraction of highly obscured AGN.

The three nearest AGN, the Circinus Galaxy, NGC~4945 and Centaurus A
are all heavily obscured (Centaurus A has a column density of $\sim$10$^{23}$
cm$^{-2}$ and an
intrinsic 2--10 keV luminosity of 6$\times10^{41}$ erg s$^{-1}$, e.g. Turner
et al. 1997; Grandi et al. 1999). 
Let us make a simple calculation to
estimate the probability to find three AGN within about 4 Mpc. Let us first 
assume that the density in the nearby Universe is representative 
of that of the Universe as a whole (the local density within a sphere of radius
corresponding to 500 km s$^{-1}$ is 1.25 times the mean density of the 
Universe, according to Schlegel et al. 1994). Integrating 
the Miyaji et al. (1998) local 0.5--2 keV 
X--ray Luminosity Function (XLF) down to 
5$\times$10$^{41}$ erg s$^{-1}$ gives a density of 2$\times$10$^{-4}$ 
Mpc$^{-3}$. This XLF is dominated by relatively unobscured AGN, with
column densities of less than 10$^{22}$ cm$^{-2}$ or so. 
The number of AGN expected from the soft XLF within 4 Mpc is then 0.05, i.e.
60 times less than observed. The probability of observing 3 or more unrelated 
AGN within that volume is then about $2\times 10^{-5}$.   
It is clear that the obscured ($>$10$^{23}$
cm$^{-2}$) sources must outnumber the unobscured ones by a large factor, 
by about an order of magnitude if the above probability is to rise to one per 
cent.

A consistency check on this 
result can be done by comparing the local Infrared Luminosity Function (ILF; 
Sanders \& Mirabel 1996) with the Piccinotti Luminosity Function (XLF;
Piccinotti et al. 1982). Heavily obscured AGN are only a minority in 
the Piccinotti sample, but they should be common in the Infrared one. 
Therefore,
the ratio between the counts in the two bands should give the maximum
ratio between obscured and unobscured AGNs. 
The two LFs have a similar shape. Applying bolometric corrections
($L_{\rm bol}\sim 10\times 
L_{\rm 2-10 keV}$ for low luminosity AGNs), it is found that the ILF is 
about 20 times the XLF. (As ULIRGs have very strong evolution, Kim
\& Sanders 1998, this number is probably even larger at higher redshift.)
Therefore, the number density of heavily obscured AGN may be as 
much as 20 times that of the 
unobscured ones, in the local Universe, and even more in the past.

The estimated ratio between obscured and unobscured sources is larger than 
usually assumed, but it must be remembered than there is increasing evidence
that many heavily obscured AGN are simply missed in optical and infrared 
surveys due
to the faintness or even lack of a (visible) Narrow Line Region (NLR: one
of the three sources just discussed, NGC~4945, is an example). 
This point is briefly discussed in the next paragraph. 

\subsection{Optical appearance}

Two sources in our sample, 
NGC~4945 and NGC~6240, are classified, based on optical emission
lines ratios, as LINERs rather than Seyfert 2 galaxies
(e.g., Veilleux \& Osterbrock 1987; Baldwin, Phillips and Terlevich 1981).
For these sources, the NLR is either
heavily obscured or absent altogether. This suggests that the 
optical classification may be misleading, not surprisingly as it is
based on properties, like the Narrow Line Regions, which after all 
are secondary and by no means necessary to {\it define} an AGN.  
(It may be worth noting that a similar warning has been given also by
Ivison et al. 2000, based on SCUBA sub-mm observations).
The fact that heavily obscured sources may lack a visible
NLR  has several possible consequences. First, estimates
of the fraction of Compton--thick sources, being based on 
[OIII] selected samples (Maiolino et al. 1998) may be biased in favour of 
less obscured sources (or sources with a smaller covering factor). 
Second, this may (at least partly) explain the
difficulties in finding type 2 QSO, when searched for in the optical. NGC~6240
is the classical example, its luminosity being well into the quasar regime.
X--ray surveys are certainly the best way of searching for highly obscured,
high luminosity sources.

\section{The iron abundance }

At 10 keV or more, where the photoelectric cut--off occurs for
Compton--thick sources, the main absorption opacity is provided by
iron, and the detailed spectrum should be dependent on the iron
abundance. In the fits described in the previous section, the iron
abundance is assumed to be the cosmic one (Anders \& Grevesse 1983).
However, the metallicity in the environment of an AGN may well be
different than in interstellar matter, and there are models of
chemical evolution in galaxies which predict a significant metallicity
enhancement (Hamann \& Ferland 1993; Matteucci \& Padovani 1993). From
a comparison between the iron line EW and the Compton reflection
continuum in a sample of Seyfert 1 galaxies observed by BeppoSAX, an
iron abundance close to the cosmic value (say, within a factor of 2 or
so) is found (Matt 2000). However, an estimate with an independent
method is certainly useful. We therefore explored the possibility to
have a significant iron overabundance in the two brightest moderately
thick sources, the Circinus Galaxy and NGC~4945.

\subsection{The Circinus Galaxy }

For the sake of simplicity, we adopted the fitting model in which
the spectrum is simply attenuated by absorption and scattering
({\sc cabs}$\star${\sc wabs} in {\sc XSPEC}),
corresponding to the physical situation of a small cloud along the line of
sight. We substituted the model {\sc varabs} to {\sc wabs}, as the former
allows for variable element abundances. We first fixed the iron abundance
to 5 times the cosmic value, finding an acceptable fit 
($\chi^{2}$=122, 119 d.o.f), but worse than the fit with the iron abundance
fixed to the cosmic value, which gives $\chi^{2}$=112/119 d.o.f. 

We then permitted the iron abundance to vary. The fit does not significantly
improve, and the best fit value for the iron abundance is, in units of the
cosmic value, equal to 1.5$\pm$0.5 (90 per cent confidence level). 
Therefore, we conclude that the 
iron abundance is consistent with the cosmic value, and constrained to be
no more than a factor of 2 larger. 

\subsection{NGC~4945}

The fit with the iron abundance assumed to be
5 times the cosmic value is significantly worse ($\chi^{2}$=94/27 d.o.f.
against $\chi^{2}$=35/28 d.o.f.), and completely unacceptable. Leaving the
iron abundance free to vary, the improvement in the quality of the fit is
marginal, and a best fit value of 1.7$^{+0.2}_{-0.6}$ is found. As in
Circinus Galaxy, also in NGC~4945 the iron abundance is constrained to be
no more than 2 times the cosmic value.

\section{The cold reflector}

The cold reflector is clearly detected in all sources but NGC~6240, 
where its presence is ambiguous, and NGC~4945, for which only an upper limit
can be obtained.  The fact that this component is so common is not 
surprising, as in Compton--thick Seyfert 2 galaxies
there is, by definition,
optically thick circumnuclear material, and unless the covering factor is 
small (but see previous section) or
the geometry particularly unfavourable, reflection from this material 
should be observed.
 
\begin{figure}[t]
\epsfig{file=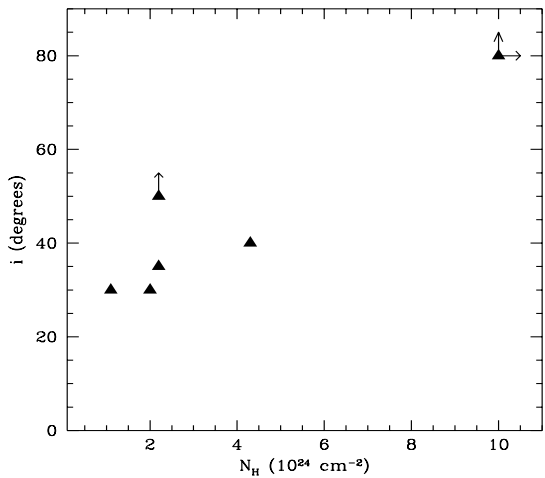, height=8cm, width=9cm}
\caption{The inclination angle of the absorbing matter as a function of 
the column density, assuming the geometry of Ghisellini, Haardt \& Matt
(1994). }
\label{nh_i}
\end{figure}

The geometry of the absorber is highly uncertain, but
let us assume the geometry envisaged by Ghisellini, Haardt \& Matt (1994), i.e. 
a torus with a half--opening angle of 30$^{\circ}$. From the ratio between
the reflected and direct luminosities it is possible to estimate the 
inclination angle of the torus, which for Circinus, NGC~6240 (from the best 
fit value; in this source the Compton reflection component is not
strongly required by the data), Mrk 3 and
Tololo~0109-383 yields a value smaller than 45$^{\circ}$. The only exception
among moderately thick sources is NGC~4945, for which the upper limit
to the reflection component translates to a lower limit of about 50$^{\circ}$
to the inclination angle.
For NGC~1068 and NGC~7674 this calculation is not possible, because we do 
not have a measurement of the direct continuum. However, from water maser
(Greenhill et al. 1997) and X--ray  (Matt et al. 1997) measurements
there are reasons to believe that NGC~1068
is observed almost edge--on. 
The relation between inclination angle and column density is plotted
in Fig.~\ref{nh_i}. Even if the uncertanties on the inclination angles
are fairly large, and the estimate admittedly model--dependent, 
there may be a correlation between the two quantities (dominated, 
it must be said,
by NGC~1068), which is naturally explained if there is a 
density profile in the torus, matter being more dense in the equatorial plane.

Finally, it is worth noting that the small inclination angles deduced
for the moderately thick sources is derived from the rather large ratio
between reflected and direct intensities. If the covering factor of the
cold matter were much smaller than in our assumed geometry, it would be very 
difficult to account for the observed values of this ratio. This
is a further argument in favour of a large covering factor of the cold
circumnuclear matter.

\section{Where does the fluorescent iron line originate?}

As discussed above, in most sources there is clear 
evidence for Compton reflection from neutral (or low ionization)
matter. It is natural to attribute the fluorescent
6.4 keV iron line to the same matter (e.g. Matt, Brandt \& Fabian 1997). 
However, at least part of the line may arise from transmission rather 
than reflection, if the column density is not too large. Of course,
it is quite possible that the absorbing and reflecting matter are just one and
the same medium. In this case, the distinction is between line photons
emitted from the inner and directly visible surface of the matter, and those
escaping from the outer boundary to reach the observer. If this is the
case, the relative importance of reflected and transmitted line photons
depends on the geometry of the system (the reflected photons increasing
with the fraction of illuminated matter observable to us)
and the covering factor of the matter. 

\begin{figure}[t]
\epsfig{file=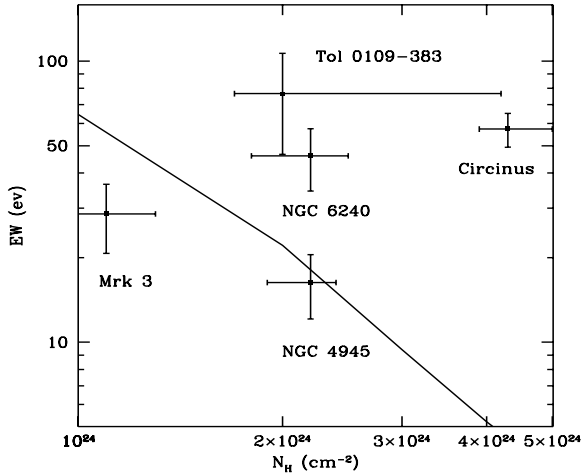, height=8cm, width=9cm}
\caption{The expected (solid line) iron line equivalent width for
a spherical distribution of matter, as a function of the column density.
The observed values are also plotted.  }
\label{ewtr}
\end{figure}

To evaluate the relative importance 
of transmission for the fluorescent iron line,
we have calculated, by means of Monte Carlo simulations, 
the line equivalent width (with respect to the continuum impinging on the
inner surface of the absorbing matter)
as a function of the column density of the matter, assumed to be
spherically distributed. The expected
values, as well as the observed ones for the sources for which the column
density can be measured, 
are shown in Fig.~\ref{ewtr}. The error bars have been calculated
considering only the statistical error on the line flux. The error on the
continuum is systematic rather than statistic, being largely due to the
uncertainty on the modeling of the absorber. For Circinus the calculated 
value falls dramatically short of the observed one, implying that
the dominant line component is the reflected one (see also Matt et al. 1999).
The same is probably true for Tololo~0109-383, but here the error bars,
especially on the column density, are too large to permit a definite
conclusion. On the contrary, 
for NGC~4945 and Mrk~3 the transmitted component may entirely account for
the observed iron line (for Mrk~3, a slight 
iron underabundance would be even 
required). While the spectral fitting of  NGC~4945 does not actually
require a Compton reflection component, for Mrk~3 it appears to be
an important ingredient; the only solution to this apparent paradox 
is that the covering factor of the matter is significantly smaller than 1. 
For NGC~6240, a reflection component
seems also required, but unfortunately the spectral fitting is rather ambiguous
in this respect.

\section{The ionized reflector(s)}

In the X--ray spectrum of 
some of the sources discussed here there is also evidence for a significant
contribution from warm (i.e. mildly ionized) and/or hot (i.e. highly ionized)
reflectors. They are best studied by looking at line emission,
and therefore BeppoSAX is not the ideal satellite in this respect, even if
the broad band has improved our knowledge of the ionized 
reflector(s) in NGC~1068 (Guainazzi et al. 1999) by better constraining its
shape and luminosity, as well as by detecting the O {\sc vii} emission line.
There is no doubt, however, that substantial progress will be achieved 
by high resolution instruments like gratings and calorimeters. 

An intriguing result is the possible detection of energy dependent
flux variability in NGC~1068 on the time scale of about a year, 
best explained in terms of a spectral variation of the ionized reflector
component (Guainazzi et al. 2000b). This, if true, implies that the size of 
the ionized reflector should be of the order of a parsec at most. A more
detailed discussion on the implications of this finding can be found
in the above paper. 


\section{Comparison with the Infrared: AGN vs. Starburst}

\begin{figure}[t]
\epsfig{file=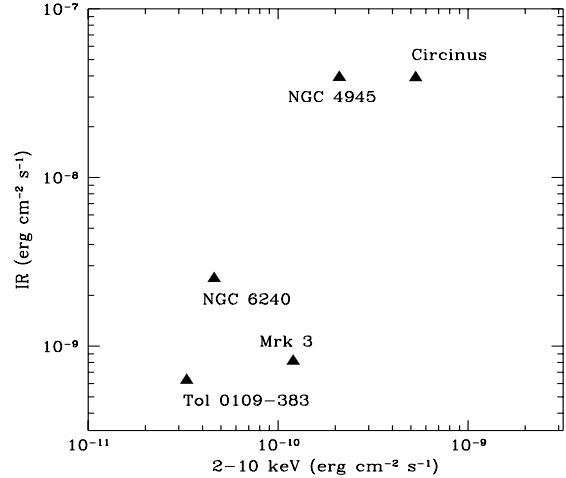, height=8cm, width=9cm}
\caption{ IR (see text) vs. 2--10 keV X--ray flux (the latter corrected
for absorption) for the 5 ``moderately" thick sources.}
\label{xir}
\end{figure}

For the 5 objects in our sample for which we have a measurement of the
intrinsic X--ray flux, we can compare it to the IR flux. For the latter, 
we used the IRAS colours as reported in the NASA/IPAC Extragalactic
Database (NED: {\sc http://nedwww.ipac.caltech.edu/}) apart from Circinus,
for which the fluxes were calculated on the basis of ISO results (Moorwood
et al. 1996; R. Maiolino and A. Marconi, private communication). 
In Fig.~\ref{xir} we show the 2--10 keV emission (corrected for absorption)
vs. the IR flux, the latter defined as: 

\begin{equation}
F_{IR}=S_{12}\nu_{12} + S_{25}\nu_{25} + S_{60}\nu_{60} + S_{100}\nu_{100} 
\end{equation}

\begin{figure}[t]
\epsfig{file=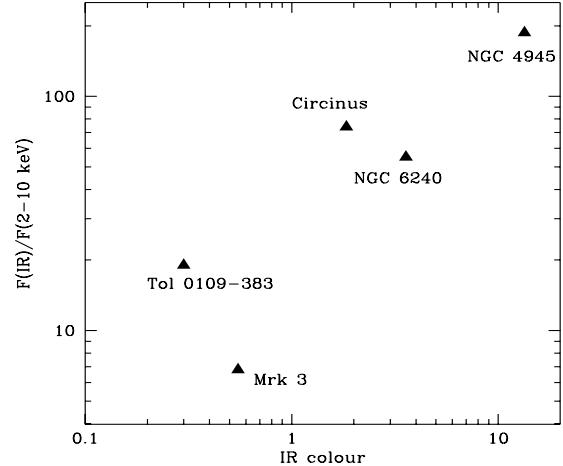, height=8cm, width=9cm}
\caption{The ratio between the IR  and 2--10 keV X--ray fluxes vs 
the IR colour, defined as: 
$(S_{60}\nu_{60} + S_{100}\nu_{100})/(S_{12}\nu_{12} + S_{25}\nu_{25})$
}
\label{colours}
\end{figure}

In Fig.~\ref{colours} we report the  IR/X--ray ratio versus an IR colour,
defined as 
$(S_{60}\nu_{60} + S_{100}\nu_{100})/(S_{12}\nu_{12} + S_{25}\nu_{25})$.
The ratio is dramatically different from source to
source, ranging from 7 for Mrk~3 to 187 for NGC~4945. As can be
seen from Fig.~\ref{colours}, this ratio is higher for the sources with 
the cooler IR colour.
A cool colour may indicate either an important
contribution from a  starburst component or strong absorption at shorter
wavelengths. The fact that the IR/X-ray ratio is larger for the cooler
sources  favours the former hypothesis. 

A related question is what is the main source of emitted power, i.e. accretion
vs. starlight. For Mrk~3 and Tololo~0109-383, both IR colours and 
IR/X--rays flux ratio
indicate unambiguously that the AGN dominates over starburst. The situation 
is more complex for the other three sources, which have cooler IR colours
suggestive of a strong starburst contribution. As discussed by Vignati et al.
(1999), for NGC~6240 the bolometric AGN luminosity derived assuming the 
Quasar SED of Elvis et al. (1994), i.e. $L_{\rm bol}\sim 30 L_{\rm 2-10 kev}$,
 is of the same order of the IR luminosity. Therefore, 
a large fraction of the IR flux should arise from the 
reprocessing of the nuclear UV and
X--ray photons, even if a significant contribution from starburst it is also
possible, especially at the longer wavelengths.

For Circinus and NGC~4945 the situation is different. 
For low-luminosity AGNs like these two Seyferts 
(which are the sources  with the lowest intrinsic $L_X$ in our
sample by far), the bolometric correction should be lower (about 10
instead of 30), and it is possible that the IR flux is actually
dominated by starburst emission (it is worth noting that the required 
starburst luminosity for Circinus and NGC~4945 is 
much lower than that would be required to account for the IR
luminosity of NGC~6240)

\section{Are there type 2 QSO?}

Finally, a few general considerations on the vexed question of the existence
of type 2 quasars may be appropriate here. At least two sources in our sample
have a 2--10 keV luminosity of order of 10$^{44}$ erg s$^{-1}$, and therefore, 
using the SED of Elvis et al. (1994), a bolometric luminosity equal or exceeding
10$^{45}$ erg s$^{-1}$, well within the Quasar regime (there are also
other examples, as listed by Vignati et al. 1999). 
But even more important is to remark that
the QSO 2 debate is largely based on a misunderstanding, as the case
of NGC~6240 makes clear. If one assumes the classical optical spectroscopic
classification,  NGC~6240 is certainly not a type 2 QSO, since it 
lacks the emission line spectrum typical of the Narrow Line Region. 
However, it must not be forgotten that the original classification, even
if it has been very useful in the past, is based on AGN properties  that
are secondary and not needed to define an AGN in the modern sense.
In X--ray terminology, a type 2 AGN is simply a source in which the 
X--ray emitting region (i.e. the black hole and its immediate surroundings)
is obscured, no matter if the BLR and NLR are visible or not.
 Since X--ray emission is a {\it 
fundamental} property of an AGN, a classification 
based on X--ray properties is clearly to be preferred, being much less 
ambiguous. Moreover, and contrary to the infrared band, in X--rays
accretion is certainly 
much more important than emission associated with stellar processes. The 
issue of the fraction and luminosity dependence of ``type 2" (or, better,
obscured) AGN will therefore be addressed (and hopefully 
settled) by X--ray surveys like those that will be performed 
by {\it Chandra} and XMM.

\section{Conclusions}

Compton--thick Seyfert 2 galaxies are very likely the most common 
subclass of AGN in the local Universe (Maiolino et al. 1998), and 
possibly also at high redshifts (Fabian 1999). The hard X--ray band is
certainly the best with which to study these sources, because
part of the nuclear radiation can penetrate the obscuring matter,
if the column density does not exceed a few times 10$^{24}$ cm$^{-2}$.
This is the reason why BeppoSAX has permitted a great advance in this
limited but important field. Unfortunately, even this instrument does not
allow the exploration of hard X--rays beyond the local Universe, and the
cosmological evolution of the column density and covering factor of the 
absorber, which are important in order to understand the growth of the 
black holes and its relation with the star formation rate (e.g. Fabian 1999),
is still unknown. Moreover, 
only a small fraction of the extragalactic sky has been covered
so far at these energies with sufficient sensitivity, 
which implies that many sources
like NGC~4945 and Circinus are still awaiting discovery. 
To make significant progresses in this field, a large 
improvement in sensitivity (like that will be provided 
by Constellation--X\footnote{\sc http://constellation.gsfc.nasa.gov/}), 
and  large area, deep surveys (like
that provided by Swift\footnote{\sc http://swift.gsfc.nasa.gov/} and, even better, that proposed with the 
EXIST\footnote{\sc http://hea-www.harvard.edu/EXIST/EXIST.html}
project) are needed.

\section*{Acknowledgments}

We thank Enzo Branchini, Roberto Maiolino and Alessandro Marconi
for useful discussions. GM acknowledges ASI and MURST (grant 
{\sc cofin}98--02--32) for financial support. ACF thanks the Royal Society
for support.

This research has made use of the NASA/IPAC Extragalactic Database (NED)
which is operated by the Jet Propulsion Laboratory, California Institute of
Technology, under contract with the National Aeronautics and Space
Administration. 

{}


\begin{thebibliography}{}

\bibitem[]{} Anders E., Grevesse N., 1989, Geochimica and Cosmochimica Acta, 
53, 197

\bibitem[]{} Baldwin J. A., Phillips M. M., Terlevich R., 1981, PASP, 93, 5

\bibitem[]{} Bianchi S., et al., 2000, in preparation

\bibitem[]{} Boella G., et al. 1997, A\&AS, 112, 299

\bibitem[]{} Cappi M., et al., 1999, A\&A, 344, 857

\bibitem[]{} Done C., Madejski G.M., Smith D.A., 1996, ApJ 463, L63

\bibitem[]{} Elvis M., et al., 1994, APJS 95, 1 

\bibitem[]{} Fabian A.C., 1999, MNRAS, 308, L39

\bibitem[]{} Fabian A.C., Iwasawa K., 1999, MNRAS, 303, L34

\bibitem[]{} Frontera F., Costa E.,  Dal Fiume D., Feroci M., Nicastro L., 
Orlandini M., Palazzi E., Zavattini G., 1997, A\&AS, 112, 357

\bibitem[]{} Ghisellini G., Haardt F., Matt G., 1994, MNRAS, 267, 743

\bibitem[]{} Grandi P., et al., 1999, Adv. Space Res, 25, 485

\bibitem[]{} Greenhill L.J., Ellingsen S.P., Norris R.P., Gough R.G., Sinclair 
M.W., Moran J.M., Mushotzky R.F., 1997, ApJ, 474, L103

\bibitem[]{} Guainazzi M., et al., 1999, MNRAS, 310, 10

\bibitem[]{} Guainazzi M., Matt G., Brandt W.N., Antonelli L.A., Barr P.,
Bassani L., 2000a, A\&A, in press (astro--ph/0001528)

\bibitem[]{} Guainazzi M., Molendi S., Vignati P., Matt G., 
Iwasawa K., 2000b, New As., in press (astro--ph/9910193)

\bibitem[]{} Hamann F., Ferland G., 1993, ApJ, 418, 11

\bibitem[]{} Ivison R., Smail I., Berger A., Kneib J.-P., Blain A., Owen F.,
Kerr T., Cowie L., 2000, MNRAS, in press (astro--ph/9911069)

\bibitem[]{} Iwasawa K.,  et al., 1993, ApJ 409, 155

\bibitem[]{} Iwasawa K., 1996, Ph.D. thesis, Nagoya University

\bibitem[]{} Iwasawa K., Fabian A.C., Matt G., 1997, MNRAS, 289, 443

\bibitem[]{} Iwasawa K., Comastri A., 1998, MNRAS, 297, 1219

\bibitem[]{} Iwasawa K., et al., 2000, in preparation

\bibitem[]{} Kim D.-C., Sanders D.B., 1998, ApJS, 119, 41

\bibitem[]{} Maiolino R., Salvati M., Bassani L., Dadina M., 
Della Ceca R., Matt G., Risaliti G., Zamorani G., 1998, A\&A, 338, 781

\bibitem[]{} Malaguti G., Palumbo G.C.C., Cappi M., Comastri A., Otani C., 
Matsuoka M., Guainazzi M., Bassani L., Frontera F., 1998, A\&A, 331, 519

\bibitem[]{} Matt G., Brandt W.N,, Fabian A.C., 1996, MNRAS, 280, 823 

\bibitem[]{} Matt G., et al., 1997, A\&A, 325, L13 

\bibitem[]{} Matt G., 2000, Astron. Lett. \& Comm., submitted

\bibitem[]{} Matt G., Pompilio F., La Franca F., 1999, New As., 4, 191

\bibitem[]{} Matt G., et al., 1999, A\&A, 341, 39

\bibitem[]{} Matteucci F., Padovani P., 1993, ApJ, 419, 485

\bibitem[]{} Miyaji T., Ishisaki Y., Ogasaka Y., Ueda Y.,  Freyberg M. J.,
Hasinger G.,  Tanaka Y., 1998, A\&A, 334, L13

\bibitem[]{} Moorwood A.F.M., Lutz D., Oliva E., Marconi A., Netzer H., Genzel 
R., Sturm E., De Graauw T., 1996, A\&A, 315, L109

\bibitem[]{} Mulchaey J.S., Koratkar A., Ward M.J., Wilson A.S., Whittle M., 
Antonucci R.R.J., Kinney A.L., Hurt T., 1994, ApJ, 436, 586

\bibitem[]{} Netzer H., Turner T.J., 1997, ApJ, 488, 694

\bibitem[]{} Netzer H., Turner T.J., George I.M., 1998, ApJ, 504, 680

\bibitem[]{} Piccinotti  G., Mushotzky R.F., Boldt E.A., Holt S.S.,
 Marshall  F.E.,  Serlemitsos P.J., Shafer R.A., 1982, ApJ, 253, 485

\bibitem[]{} Risaliti G., Maiolino R., Salvati M., 1999, ApJ, 522, 157

\bibitem[]{} Sanders D.B., Mirabel I.F., 1996, ARAA, 34, 749

\bibitem[]{} Schlegel D., Davis M., Summers F., Holtzman J.A., 1994, ApJ, 
427, 527

\bibitem[]{} Turner T.J., George I.M., Mushotzky R.F., Nandra K., 1997, ApJ,
475, 118

\bibitem[]{} Veilleux S., Osterbrock D. E., 1987, ApJS, 63, 295

\bibitem[]{} Vignati P., et al., 1999, A\&A, 349, L57

\end{thebibliography}
\end{document}